\documentclass[letter]{aa}
\usepackage{graphicx}
\usepackage{natbib}
\usepackage{txfonts}

\begin{document}

   \title{Evidence of an age gradient along the line of sight in the nuclear stellar disc of the Milky Way}

  \author{F. Nogueras-Lara
          \inst{1}       
           \and      
          M. Schultheis
          \inst{2}       
          \and
          F. Najarro 
          \inst{3}    
          \and
          M. C. Sormani
          \inst{4} 
          \and
          D. A. Gadotti
          \inst{5}           
          \and
          R. M. Rich 
          \inst{6}                         
          }

   \institute{
    Max-Planck Institute for Astronomy, K\"onigstuhl 17, 69117 Heidelberg, Germany
              \email{francisco.nogueraslara@eso.org}       
              \and 
        Université Côte d'Azur, Observatoire de la Côte d'Azur, Laboratoire Lagrange, CNRS, Blvd de l'Observatoire, F-06304 Nice, France           \and 
     Centro de Astrobiolog\'ia (CSIC/INTA), ctra. de Ajalvir km. 4, 28850 Torrej\'on de Ardoz, Madrid, Spain     
     \and
     Zentrum für Astronomie, Institut für theoretische Astrophysik, Universität Heidelberg, Albert-Ueberle-Str 2, D-69120 Heidelberg, Germany    
     \and
     Centre for Extragalactic Astronomy, Department of Physics, Durham University, South Road, Durham DH1 3LE, UK
           \and
    Department of Physics and Astronomy, UCLA, 430 Portola Plaza, Box 951547, Los Angeles, CA 90095-1547                                
       }
   \date{}

 
  \abstract
   {The nuclear stellar disc (NSD) is a flat dense stellar structure at the heart of the Milky Way. Recent work has shown that analogous structures are common in the nuclei of external spiral galaxies, where there is evidence of an age gradient that indicates that they form inside-out. However, the characterisation of the age of the NSD stellar population along the line of sight is still missing due to its extreme source crowding and the high interstellar extinction towards the Galactic centre.}
   {We aim to characterise the age of the stellar population at different average Galactocentric NSD radii to investigate for the first time the presence of an age gradient along the line of sight.} 
    {We selected two groups of stars at different NSD radii via their different extinction and proper motion distribution. We analysed their stellar population by fitting their de-reddened $K_s$ luminosity functions with a linear combination of theoretical models.}
   {We find significant differences in the stellar population at different NSD radii, indicating the presence of an age gradient along the line of sight. Our sample from the closest edge of the NSD contains a significant fraction ($\sim40$\,\% of its total stellar mass) of intermediate-age stars (2-7\,Gyr) that is not present in the sample from stars deeper inside the NSD, in which $\sim90\,\%$ of the stellar mass is older than 7\,Gyr. Our results suggest that the NSD age distribution is similar to the one found in external galaxies and they  imply that bar-driven processes observed in external galaxies are similarly at play in the Milky Way.}  
   {}

\titlerunning{Evidence of an age gradient along the line of sight in the NSD}
\authorrunning{F. Nogueras-Lara et al.}
   \maketitle
%

\section{Introduction}

The nuclear stellar disc (NSD) is a flat dense stellar structure that roughly outlines the Galactic centre with a radius of $\sim150$\,pc and a scale height of $\sim40$\,pc \citep[e.g.][]{Launhardt:2002nx,gallego-cano2019,Sormani:2020aa,Sormani:2022wv}. Its study is hampered by the extreme source crowding and the high extinction towards the innermost regions of the Galaxy \citep[e.g.][]{Nishiyama:2006tx,Nogueras-Lara:2021wj}.The analysis of its structure and stellar population is therefore mainly limited to near-infrared high-resolution photometry, while spectroscopy is restricted to very small regions of high interest \citep[e.g.][]{Lohr:2018aa,Clark:2018aa}.

A photometric analysis of the innermost $\sim1600$\,pc$^2$ of the NSD using the GALACTICNUCLEUS catalogue \citep[a high-angular, $\sim0.2''$, resolution photometric survey in the near-infrared,][]{Nogueras-Lara:2018aa,Nogueras-Lara:2019aa,Nogueras-Lara:2019ad}, determined that the NSD is mainly dominated by old stars (more than 80\,\% of its total stellar mass was formed more than 8\,Gyr ago). A similar analysis of Sgr\,B1, a region of intense HII emission located $\sim$100\,pc away from the centre of the NSD \citep[e.g.][]{Simpson:2021ti}, showed evidence of an intermediate-age stellar population ($\sim40$\,\% of the stellar mass was formed $\sim2-7$\,Gyr ago) in addition to the old stars \citep[$\sim40\,\%$ of the stellar mass is older than 7\,Gyr,][]{Nogueras-Lara:2022ua}. This is consistent with the inside-out formation scenario proposed for the NSD \citep[based on recent studies on nearby spiral galaxies,][]{Gadotti:2020aa,Bittner:2020aa} that would originate a radial age gradient. In this picture, the formation of the NSD is related to the Galactic bar, which funnels gas from the Galactic disc towards the innermost regions of the Galaxy \citep[e.g.][]{Sormani:2019aa}, producing an accumulation of gas and dust that grows the NSD from inside-out. 

In this Letter, we follow up on previous work that studied stellar populations at different lines of sight \citep{Nogueras-Lara:2022ua} and investigate the age distribution of stellar populations located at different NSD radii along the line of sight. We aim to assess the presence of an age gradient that would support the inside-out formation scenario proposed for the NSD.

\section{Data}

\subsection{Photometry}
\label{satu}

              \begin{figure*}[h!]
   \includegraphics[width=\linewidth]{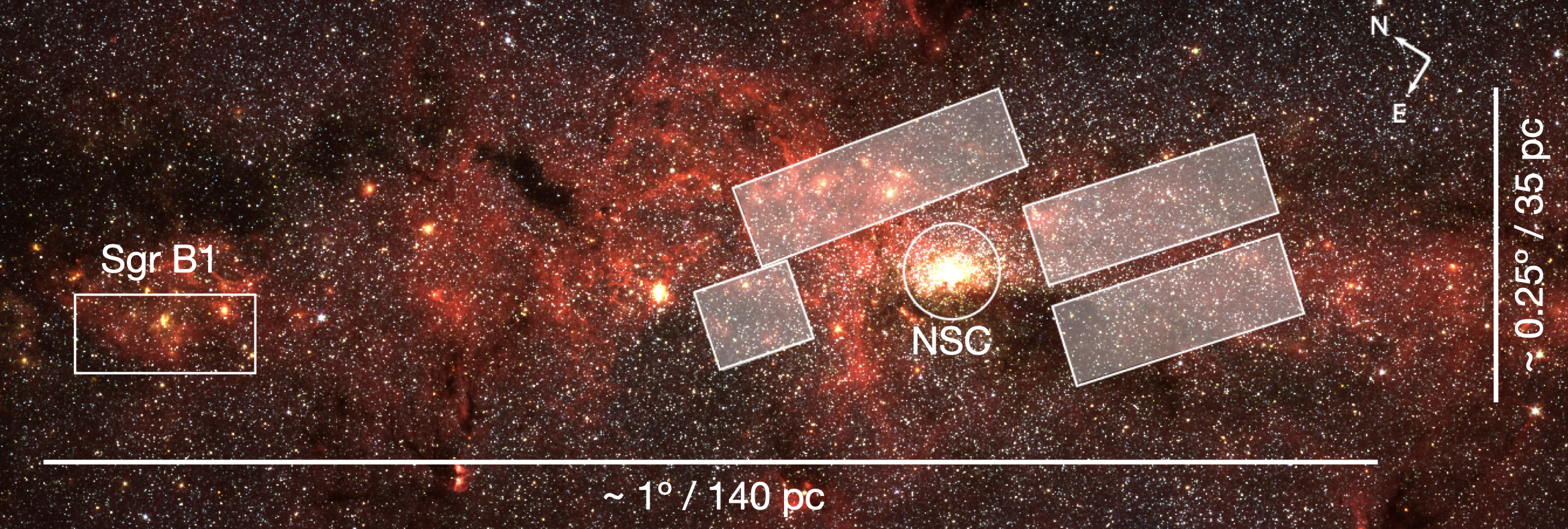}
   \caption{Spitzer false colour image using 3.6, 4.5, and 5.8\,$\mu$m, as red, green, and blue, respectively \citep{Stolovy:2006fk}. The white shaded boxes indicate the target region. The position of Sgr\,B1 as well as the nuclear star cluster (NSC) are indicated. The circular region indicates the effective radius of the nuclear star cluster \citep[$\sim5$\,pc, e.g.][]{gallego-cano2019}.}

   \label{scheme}
    \end{figure*}

We used $H$ and $K_s$ photometry from the GALACTICNUCLEUS survey \citep{Nogueras-Lara:2018aa,Nogueras-Lara:2019aa}. This is a high-angular ($\sim0.2''$) resolution near-infrared catalogue specially designed to observe the NSD. It contains accurate point spread function photometry for more than three million sources. The zero point systematic uncertainty is below 0.04\,mag in all bands, whereas the statistical uncertainties are below 0.05\,mag at $H\sim19$\,mag and $K_s\sim18$\,mag. 

We corrected potentially saturated sources in $K_s$ \citep[$K_s<11.5$\,mag, e.g.][]{Nogueras-Lara:2019aa} using $K_s$ data from the SIRIUS IRSF survey \citep{Nishiyama:2008qa}. We also included saturated sources which were not detected in the GALACTICNUCLEUS catalogue.

Figure\,\ref{scheme} indicates the target region, which was chosen to overlap with a high-precision proper motion catalogue of the Galactic centre \citep{Libralato:2021td}. We excluded regions close to the nuclear star cluster because its different star formation history \citep[e.g.][]{Schodel:2020aa,Nogueras-Lara:2021wm} could affect our analysis.


\subsection{Proper motions}

We used the Galactic centre proper motion catalogue obtained by \citet{Libralato:2021td} using the Wide-Field Camera 3 (WFC3/IR, filter F153M) at the HST. This catalogue contains absolute high-precision proper motions for more than 800,000 stars calibrated with Gaia DR2 \citep{Gaia-Collaboration:2016uw,Gaia-Collaboration:2018aa}.

\section{Colour-magnitude diagram and targets selection}
\label{CMD_sect}

\begin{figure}
   \includegraphics[width=\linewidth]{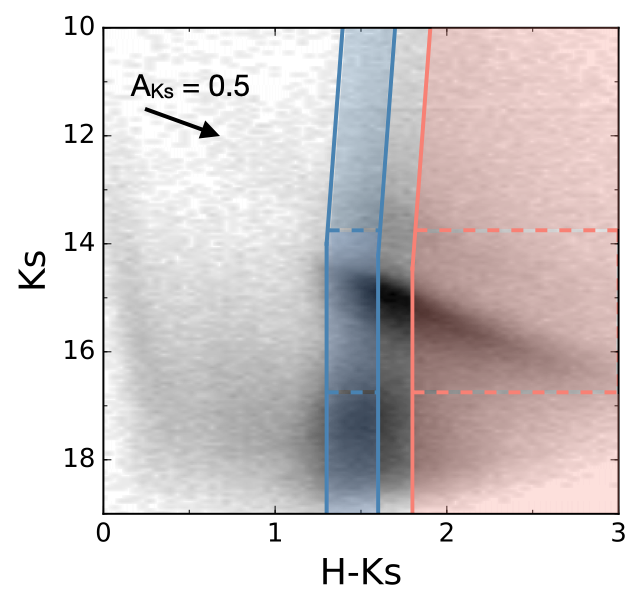}
   \caption{CMD $K_s$ versus $H-K_s$. The blue and salmon coloured regions indicate the two target groups of stars with different reddening dominated on average by stars from the closest edge of the NSD (NSD outer region) and stars deeper inside the NSD  (NSD inner region), respectively. The dashed boxes show the reference stars used to build the extinction maps for each of the stellar groups analysed (see Sect.\,\ref{extinct}). The black arrow indicates the reddening vector.}

   \label{CMD}
    \end{figure}

Figure\,\ref{CMD} shows the colour magnitude diagram (CMD) $K_s$ versus $H-K_s$ of the target region. To remove foreground stars from the Galactic disc and bulge/bar, we applied a colour cut $H-K_s\gtrsim1.3$\,mag \citep[e.g.][]{Nogueras-Lara:2021uz,Nogueras-Lara:2021wj}.

Recent results on the NSD structure indicate that the extinction within the NSD correlates with the position of the stars along the line of sight \citep{Nogueras-Lara:2022aa}. Hence, stars belonging to the closest edge of the NSD suffer a lower reddening than stars from its farthest edge. It is then possible to statistically distinguish between stars located at different NSD radii along the line of sight by applying a colour cut in the CMD $K_s$ versus $H-K_s$. We selected two target groups corresponding to stars from the closest edge of the NSD and stars deeper inside the NSD (hereafter NSD outer and inner regions, respectively) by applying the colour cuts indicated in Fig.\,\ref{CMD}.

\begin{figure}
   \includegraphics[width=\linewidth]{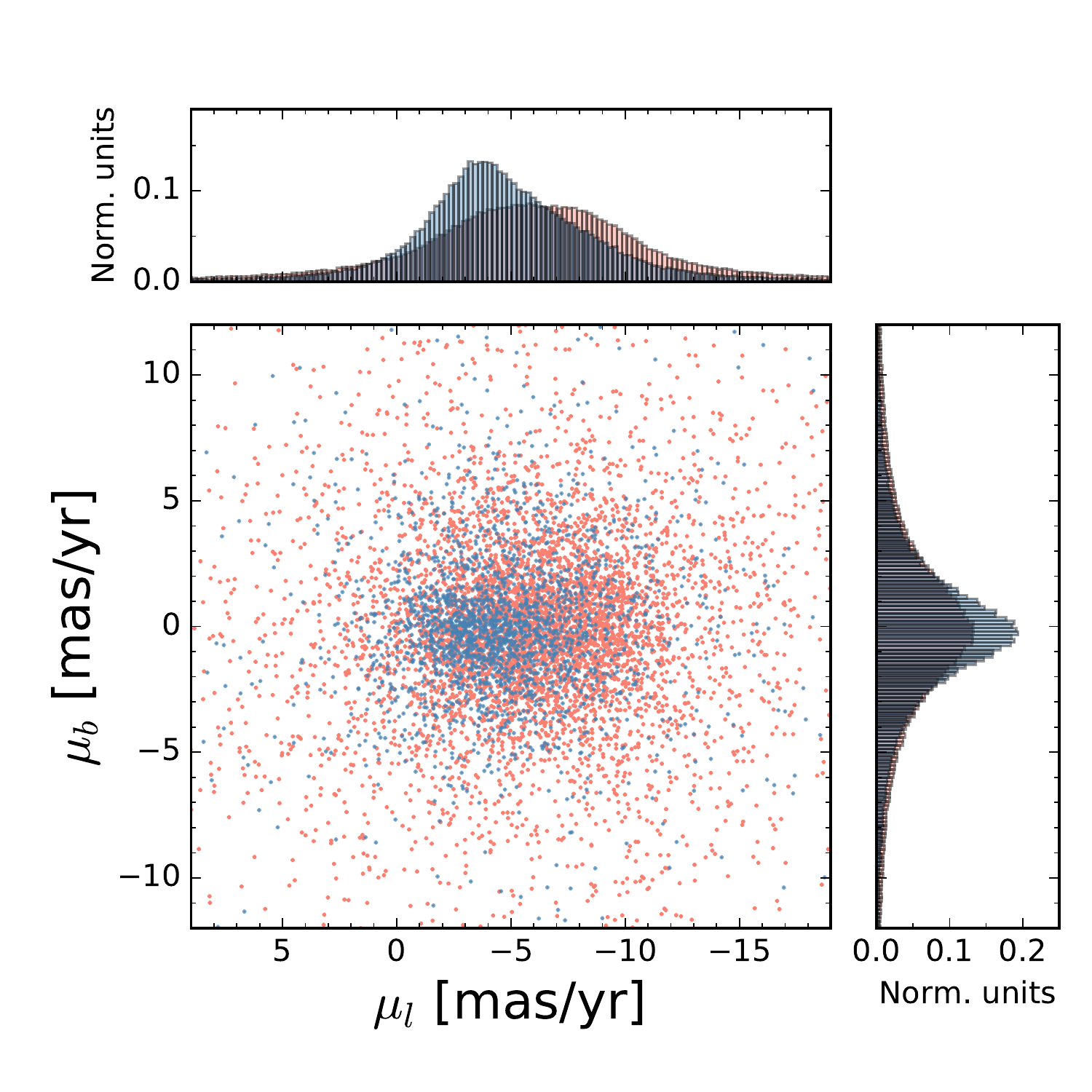}
   \caption{Proper motion distribution of stars from each of the target groups. The blue and salmon dots indicate stars from the NSD outer and inner regions, respectively. Only a fraction of the stars is shown to not overcrowd the plot.}

   \label{proper}
    \end{figure}

To check our target selection and assess whether the target stellar groups are dominated by stars located at different NSD radii, we cross-correlated the photometry from the GALACTICNUCLEUS survey \citep{Nogueras-Lara:2018aa,Nogueras-Lara:2019aa} with the proper motion catalogue by \cite{Libralato:2021td}. Figure\,\ref{proper} shows the proper motion distribution of stars from each of the groups. We clearly distinguish a different distribution of the proper motion component parallel to the Galactic plane ($\mu_l$) for each of the groups. We computed a mean value of $\mu_{l} = -4.66\pm0.02$\,mas/yr for the group of stars with the bluest colour (i.e. the lowest extinction), where the uncertainty corresponds to the standard error of the distribution. This value is in agreement with the rotation of stars from the closest edge of the NSD, $\mu_l = -4.70\pm0.03$\,mas/yr, obtained in \citet{Nogueras-Lara:2022aa}. On the other hand, we obtained a mean value of $\mu_{l} = -5.77\pm0.03$\,mas/yr for the group of stars with the reddest colour (i.e. the highest extinction). This value is lower than the one computed in previous work for stars from the farthest edge of the NSD \citep[$\mu_{l} = -7.65\pm0.03$\,mas/yr, see Table\,1 in ][]{Nogueras-Lara:2022aa}, but significantly larger than the result for stars from the closest edge. This indicates that the stars in this group mainly belong to the inner regions of the NSD, being placed at more internal radii than stars from the farthest edge of the NSD and the outer group previously defined.

We found that the proper motion distribution of the component perpendicular to the Galactic plane is $<\mu_b>\sim 0 $\,mas/yr, which is in agreement with previous work \citep{Shahzamanian:2021wu,Martinez-Arranz:2022uf,Nogueras-Lara:2022aa}. Small differences in the shape of the $\mu_b$ distribution of the target groups might be caused by a different contribution of contaminant stars from the Galactic bulge/bar which can depend on the applied colour cut for the sample selection.

\section{Stellar population analysis}
     
To analyse the stellar population in each of the target groups, we created $K_s$ luminosity functions and fitted them with a linear combination of theoretical models following the method in \citet{Nogueras-Lara:2019ad}, \citet{Schodel:2020aa}, and \citet{Nogueras-Lara:2022ua}. 

\subsection{Reddening correction}
\label{extinct}

We de-reddened the stars in each of the target groups by creating dedicated extinction maps following the approach described in \citet{Nogueras-Lara:2021wj} and using the extinction curve $A_H/A_{K_s} = 1.84\pm0.03$ \citep{Nogueras-Lara:2020aa}. To build the extinction maps, we used red clump stars \citep[giant stars in their helium core-burning sequence, e.g.][]{Girardi:2016fk} and  other red giant stars with a similar brightness (dashed boxes in Fig.\,\ref{CMD}), as reference stars, and assumed that they have an intrinsic colour of $(H-K_s)_0=0.10\pm0.01$ \citep{Nogueras-Lara:2021wj}. We defined a pixel size of $\sim10''$ and used the five closest reference stars within a maximum radius of $15''$ from the centre of each pixel to compute the extinction value applying an inverse distance weight method \citep[for details see appendix A.2 of ][]{Nogueras-Lara:2021wj}. We only computed an extinction value for a given pixel if at least five reference stars were found within the maximum radius.

\begin{figure}
   \includegraphics[width=\linewidth]{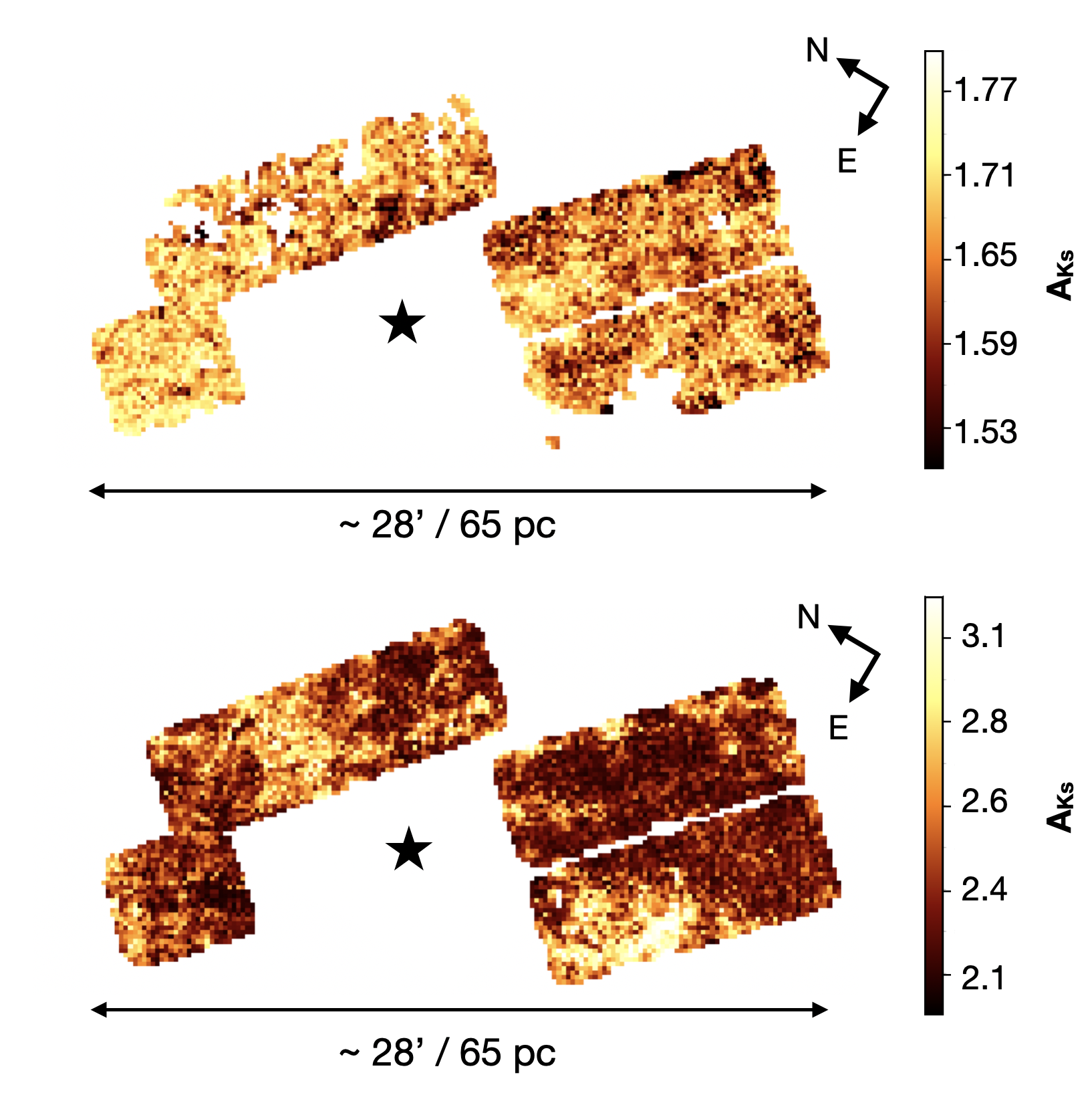}
   \caption{Extinction maps obtained for the stars from the NSD outer (upper panel) and inner regions (lower panel). The black star indicates the position of the supermassive black hole at the centre of the Galaxy. The white pixels indicate that there is no associated extinction value due to the lack of reference stars.}

   \label{ext_maps}
    \end{figure}

Figure\,\ref{ext_maps} shows the obtained extinction maps. We computed mean extinction values of $A_{K_s}\sim1.7$\,mag and $A_{K_s}\sim2.4$\,mag for the extinction maps corresponding to stars from the NSD outer and inner regions, respectively. We applied the extinction maps to de-redden the photometry of each of the target groups.

\subsection{Luminosity function}

We created a $K_s$ luminosity function using the de-reddened photometry for each of the target groups. We only used stars within the defined colour cut in Fig.\,\ref{CMD} for the group of stars from the NSD outer region, whereas we also included stars that were detected in $K_s$ but not in $H$ for the group of stars from the inner region of the NSD for the sake of completeness. Not detecting these stars in $H$ indicates that they are affected by a larger extinction and then belong to the target group of stars.

To avoid over-de-reddened stars, we excluded stars with a de-reddened $H-K_s$ colour more than 2$\sigma$ bluer than the mean value of the de-reddened distribution of the red clump features \citep[e.g.][]{Nogueras-Lara:2019ad,Nogueras-Lara:2022ua}. To build the luminosity functions, we chose the bin width that maximises the Freedman-Diaconis \citep{Freedman1981} and Sturges \citep{doi:10.1080/01621459.1926.10502161} estimators, using the 'auto' option in the python function numpy.histogram \citep{Harris:2020aa}. We assumed Poisson uncertainties (i.e. the square root of the number of stars) for each luminosity function bin.

The faint end of the luminosity functions is significantly affected by incompleteness due to the extreme source crowding in the NSD \citep[e.g.][]{Nogueras-Lara:2019ad,Nogueras-Lara:2022ua}. In this way, we computed a crowding completeness solution determining the critical distance at which a star can be detected around a brighter star, following the method described in \citet{Eisenhauer:1998tg}. We divided the analysed field into small subregions of $2'\times1.4'$ and computed a completeness solution for each of them. We then combined the results and computed their mean and standard deviation to obtain the final completeness and its uncertainty, as explained in \citet{Nogueras-Lara:2020aa}. We used our completeness solution to correct the luminosity functions, setting a lower limit of $70\,$\% of data completeness. On the other hand, due to possible saturation and incompleteness of the bright end of the luminosity function, even after using the SIRIUS IRSF catalogue to correct saturated stars (see Sect.\,\ref{satu}), we restricted the bright end of the luminosity function to $K_s=8.5$\,mag for the subsequent analysis \citep[e.g.][]{Nogueras-Lara:2022ua}.

\subsection{Stellar populations}

Luminosity functions contain fundamental information about the star formation history. In this way, the presence of particular features such as the asymptotic giant branch bump, the red clump, or the red giant branch bump, and their relative weights, allows us to characterise their stellar population \citep[e.g.][]{Nogueras-Lara:2018ab,Nogueras-Lara:2019ad,Schodel:2020aa}.

We fitted the obtained $K_s$ luminosity functions with a linear combination of theoretical models to determine the stellar populations present in each target group, as explained in \citet{Nogueras-Lara:2019ad,Nogueras-Lara:2022ua}. We used 14 Parsec models\footnote{generated by CMD 3.6 (http://stev.oapd.inaf.it/cmd)} \citep{Bressan:2012aa,Chen:2014aa,Chen:2015aa,Tang:2014aa,Marigo:2017aa,Pastorelli:2019aa,Pastorelli:2020wz}, choosing ages that homogeneously sample the possible stellar populations in the analysed luminosity functions (14, 11, 8, 6, 3, 1.5, 0.6, 0.4, 0.2, 0.1, 0.04, 0.02, 0.01, and 0.005 Gyr). The age-spacing between the models was selected to account for the differences that appear in the luminosity function for stellar populations with different ages. In this way, we increased the number of models towards younger ages, where the luminosity function is more affected by age variations \citep[see Sect. Methods in][]{Nogueras-Lara:2022ua}. 

\begin{figure}
   \includegraphics[width=\linewidth]{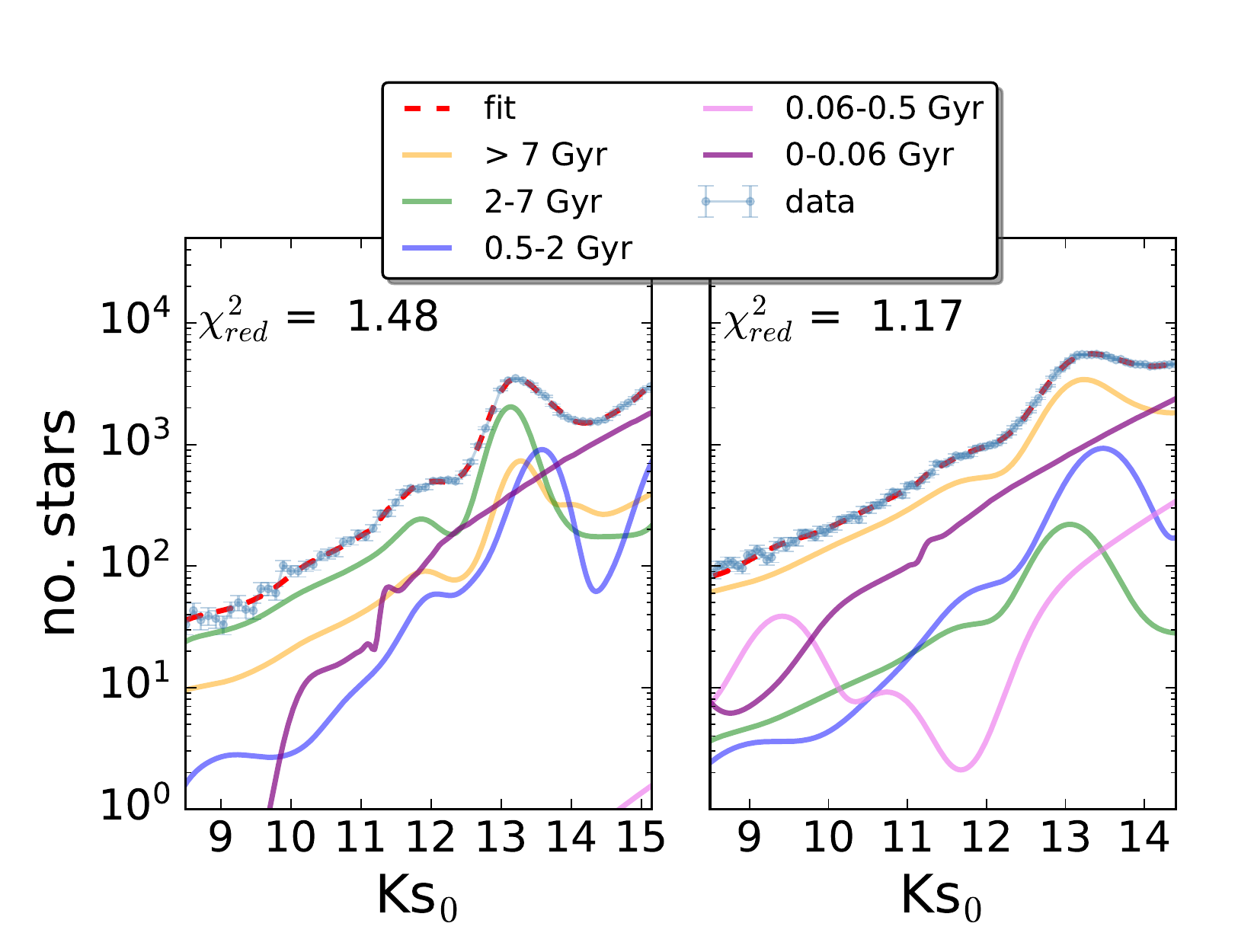}
   \caption{Analysis of the de-reddened luminosity functions corresponding to the NSD outer (left panel) and inner (right panel) regions. The reduced $\chi^2$ of the fit is shown in each panel. The 14 theoretical models used are grouped into five broader age bins to decrease the degeneracy between models with similar ages.}

   \label{KLFs}
    \end{figure}

We assumed a Kroupa initial mass function corrected for unresolved binaries \citep{Kroupa:2013wn}, and twice solar metallicity (Z=0.03) for our models, in agreement with previous results for the NSD \citep[e.g.][]{Schultheis:2019aa,Nogueras-Lara:2019ad,Schultheis:2021wf,Nogueras-Lara:2022ua}. Our fitting algorithm included a parameter to consider the distance towards the target stellar populations (including potential offsets introduced in the de-reddening process), and also a Gaussian smoothing factor accounting for possible distance differences between stars and also some residual uncorrected differential extinction. Figure\,\ref{KLFs} shows the best fits obtained by minimising a $\chi^2$ for each of the luminosity functions, where we combined models with similar ages to minimise the degeneracy.

We resorted to Monte Carlo simulations to determine the contribution of each model to the luminosity function and estimate the uncertainties, as explained in \citet{Nogueras-Lara:2022ua}. We created 1000 Monte Carlo samples by computing the number of stars per bin in each luminosity function by randomly varying the original value assuming a Gaussian distribution with a standard deviation equal to each bin uncertainty. We then fitted each Monte Carlo sample following our method and average over the results to compute a mean value and its standard deviation. 

\begin{figure*}
   \includegraphics[width=\linewidth]{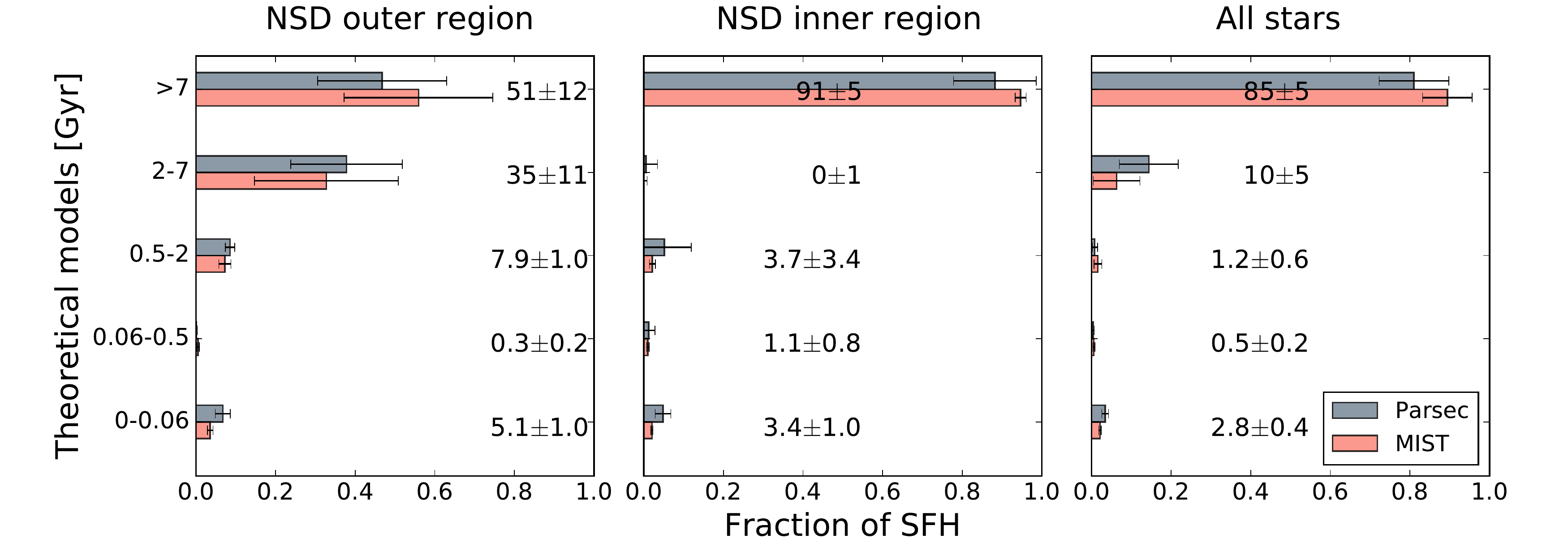}
   \caption{Stellar populations present in the NSD outer region (left panel), inner region (central panel), and all the NSD stars in the sample (right panel). The numbers indicate the percentage of mass due to a given age bin and its standard deviation.}

   \label{SFH}
    \end{figure*}

We repeated the process using also MIST models \citep{Paxton:2013aa,Dotter:2016aa,Choi:2016aa} with similar ages and a Salpeter initial mass function, to account for possible systematic effects caused by the choice of different stellar evolutionary tracks \citep[e.g.][]{Nogueras-Lara:2022ua}. We obtained small differences between the analyses carried out with Parsec and MIST models, as shown in Fig.\,\ref{SFH}. The final results were computed by combining the values obtained when using both models (Fig.\,\ref{SFH}). The uncertainties were calculated quadratically propagating the ones obtained for each age bin with Parsec and MIST models. 

From Fig.\,\ref{SFH}, we conclude that the stellar populations in each of the target regions are significantly different. While most stars across the NSD have ages above 7\,Gyr, we find that in the outer region of the NSD there is a factor 3 more stars with ages between 2 and 7\,Gyr than in its inner region, strongly suggesting the presence of an age gradient.

\subsection{Systematic uncertainties}

To assess our results, we studied potential sources of systematic uncertainties:
\begin{enumerate}

\item {\it Luminosity function bin width}. We repeated the described process for each of the target populations assuming half and double the previously computed bin width. We did not observe any significant difference within the uncertainties.

\item {\it Completeness solution and faint end of the luminosity function}. We created two additional completeness solutions considering the completeness values per magnitude bin, subtracting and adding the corresponding uncertainty. We corrected the luminosity functions applying these two completeness solutions and repeated the fitting process. We slightly changed the faint end of the luminosity function to avoid problems related to incompleteness due to sensitivity. We did not observe any significant variation within the obtained uncertainties. We also repeated the process assuming a completeness limit of 75\,\% instead of 70\,\% without any significant change.

\item {\it Bright end of the luminosity function}. We set a bright limit of $K_s=8.5$\,mag (de-reddened) for our study. Nevertheless, we also repeated our analysis including stars up to $K_s=8$\,mag and conclude that it does not affect our results in any significant way. Only a somewhat larger contribution ($\lesssim2$\,\% variations) is observed for some young age bins (<2\,Gyr) in the sample from the NSD outer region.

\item {\it Metallicity of the stellar population}. We assumed twice solar metallicity for our analysis in agreement with the super solar metallicities found for NSD stars \citep[e.g.][]{Schultheis:2021wf,Nogueras-Lara:2022tp,Feldmeier-Krause:2022vm}. Nevertheless, we also explored solar metallicity and 1.5 solar metallicity using Parsec models. In both cases we determined that the old contribution is shifted towards the age bin 2-7\,Gyr for the NSD outer region. On the other hand, there is not any variation in the results obtained for the NSD inner region. This result still agrees with the presence of different stellar populations for different NSD radii and is also compatible with an inside-out formation of the NSD. However, we determined that the reduced $\chi^2$ is larger for both target populations when considering solar or 1.5 solar metallicity, indicating that assuming twice solar metallicity gives a better result. 

\item {\it Influence of the extinction maps}. We repeated the process for each target group assuming different parameters to build the extinction map (seven reference stars, and a maximum radius of $20''$ to select reference stars). We did not observe any significant change within the uncertainties.

\item {\it Width of the colour cut for target selection}. We selected stars in the NSD outer region by applying a colour cut width $\Delta(H-K_s)=0.3$\,mag, whereas the width of the colour cut was larger for the NSD inner region (Fig.\,\ref{CMD}). Our choice was motivated by the lower completeness when considering the stars from the inner regions of the NSD. To assess our results, we also tested the effect of assuming a colour cut with a similar width ($H-K_s=0.3$\,mag) for stars in the NSD inner region. This implies that the maximum variation of the differential extinction is the same for both stellar samples. We determined that the contribution of stars older than $7$\,Gyr is $\sim80$\,\%, whereas the contribution from the age bin $2-7$\,Gyr is $\sim10$\,\%. This is probably because this stellar sample contains fewer stars from the innermost regions of the NSD than the original one. Therefore, our results are compatible with an age gradient in which stars in the colour range $H-K_s\sim 1.8-2.1$\,mag correspond to an intermediate case in between the previously analysed ones (Fig.\,\ref{SFH}.)


\item {\it Initial mass function}. Recent results on the known young clusters in the NSD point towards a top-heavy initial mass function \citep[e.g.][]{Hosek:2019va}. Nevertheless, our stellar samples are dominated by giant stars with a mass of $\gtrsim1$\,M$_\odot$ and therefore assuming a top-heavy initial mass function does not have any observable effect on our results \citep[e.g.][]{Schodel:2020aa}.

\item {\it Contamination from the Galactic bulge/bar in the target groups}. We used a colour cut $H-K_s\gtrsim1.3$\,mag to remove foreground stars from the NSD sample. However, residual contamination from the innermost regions of the Galactic bulge/bar could remain in our target groups \citep[$\lesssim20$\,\% of the stars in the sample, as estimated by ][]{Sormani:2022wv}. Given that the stellar population from the innermost bulge/bar is mainly old and metal rich \citep[e.g.][]{Clarkson:2011ys,Nogueras-Lara:2018ab,Renzini:2018aa}, its presence would not affect the detected radial age gradient, and could only increase the number of stars in the oldest age bin, without affecting our conclusions.

\item {\it Contamination between the NSD outer and inner regions}. We chose our target stellar groups statistically distinguishing between stars located at different NSD radii using their different extinctions. Nevertheless, it is not possible to completely separate stars from the outer and the inner regions of the NSD given their proximity \citep[the radius of the NSD is $\sim150$\,pc,][]{Nogueras-Lara:2022aa} and their small kinematic difference \citep[$\sim3-4$\,mas/yr between the NSD edges, e.g.][]{Shahzamanian:2021wu,Nogueras-Lara:2022aa}. To estimate the contamination between the target groups, we chose stars with well-defined proper motions  (i.e. uncertainty below 0.5\,mas/yr). For the NSD outer region, we assumed that stars with $\mu_l<-5.77+3\cdot0.02$ are contaminant stars from the inner region, where the reference value corresponds to the mean $\mu_l$ for the inner region plus 3$\sigma$ (see Sect.\,\ref{CMD_sect}). Analogously, for the NSD inner region, we estimated the fraction of stars with $\mu_l>-4.66-3\cdot0.03$ and considered that they likely belong to the outer region. We ended up with a contamination of $\lesssim30$\,\% for each of the target groups from the other one. Nevertheless, this is probably an overestimation given the low number of stars with proper motion uncertainties below 0.5\,mas/yr ($\sim15\,\%$ of the stars with proper motions). This is particularly clear for the inner NSD, in which having 30\,\% of stars from the outer NSD would imply a much larger stellar population in the 2-7\,Gyr age bin than the detected one. Actually, an alternative way to estimate the contamination from the outer NSD in the inner NSD sample is to assume that the detected stellar contribution from stars in the age bin 2-7\,Gyr is completely due to a stellar population as the one measured for the outer NSD. In this way we conclude that $\sim5$\,\% of contamination from the outer NSD stellar population can account for the estimated $\sim1$\,\% uncertainty in the age bin 2-7\,Gyr (middle panel Fig.\,\ref{SFH}). 

In any case, in spite of the potential contamination between the target populations, we determined that their stellar populations are significantly different.

\end{enumerate}

\section{Discussion and conclusion}
   
 A previous study of the central regions of the NSD (the innermost $\sim20$\,pc\,$\times\,90$\,pc excluding the nuclear star cluster)  showed that it is dominated by old stars \citep[more than 80\,\% of the total stellar mass was found to be older than 8\,Gyr,][]{Nogueras-Lara:2019ad}. In this Letter, we have analysed a similarly concentrated region towards the centre of the NSD, but considered stars located at different NSD radii along the line of sight. We selected our target samples of stars by using their different extinction, and found that they are placed on average at different Galactocentric NSD radii by analysing their proper motion distribution. Hence, the current study was only possible after \citet{Nogueras-Lara:2022aa} found that the extinction within the NSD correlates with the position of the stars along the line of sight.

Our results indicate that the stellar population at different NSD radii along the line of sight is significantly different. We detected an age gradient in the NSD along the line of sight, so that the stellar population from the innermost regions is significantly older than stars from the outer regions of the NSD. In this way, we determined that there is a significant fraction of stars with ages between 2-7\,Gyr in the NSD outer region that is not present in our sample deeper inside the NSD, where $\sim90\,$\% of the originally formed stellar mass is older than 7\,Gyr. We also observed that the contribution of even younger stars (age bins 0.5-2\,Gyr and 0-0.06\,Gyr) is also larger in the NSD outer region in comparison to the inner one.

To compare our results with the predominantly old stellar population detected in the central region of the NSD \citep{Nogueras-Lara:2019ad}, we also studied the $K_s$ luminosity function derived from all the stars in the analysed region. Figure\,\ref{SFH} (right panel) shows the result. We found that the stellar mass is dominated by old stars (>7\,Gyr) in agreement with previous studies \citep{Nogueras-Lara:2019ad,Nogueras-Lara:2022ua}. We detected a somewhat lower contribution of the age bins between 0.06-2\,Gyr in comparison with \citet{Nogueras-Lara:2019ad}, which can be explained due to the different regions analysed, the different models used, and the different age bins. The contribution from the intermediate age stellar population (2-7\,Gyr) is not significant when analysing all the stars in the region because the majority of stars in our sample belong to the innermost regions of the NSD. Therefore, the stars from the closest NSD edge do not contribute to the total luminosity function much. To further assess this, we computed the stellar mass from the NSD outer region in the age bin $2-7$\,Gyr, using Parsec models \citep[see Sect. Methods in][]{Nogueras-Lara:2022ua}. We ended up with a total mass of $(5.3\pm1.8)\cdot10^6$\,M$_\odot$ in the age bin $2-7$\,Gyr. Comparing this value with the total stellar mass obtained when using all the stars in the target region, $(1.3\pm0.1)\cdot10^8$\,M$_\odot$, we conclude that the mass of stars in the age bin $2-7$\,Gyr only accounts for $\sim4$\,\% of the total stellar mass, which is compatible with the results in Fig.\,\ref{SFH} (right panel), and also with previous work, where the old stellar component dominates \citep{Nogueras-Lara:2019ad,Nogueras-Lara:2022ua}. This is in agreement with an exponential radial scale length of the NSD, in which the stellar mass  density increases towards its innermost regions \citep[e.g.][]{Sormani:2022wv}.

Our results, in combination with the detection of a significant stellar mass in the age bin of 2-7\,Gyr in the Sgr\,B1 region (belonging to the NSD and located at $\sim100$\,pc in projection from its centre), indicate the presence of an age gradient in the NSD. Therefore, the age distribution of the NSD stellar population is similar to the one found for other NSDs in external spiral galaxies \citep{Bittner:2020aa}, and it supports an inside-out formation of the NSD \citep{Gadotti:2020aa,Bittner:2020aa}.

  \begin{acknowledgements}
This work is based on observations made with ESO Telescopes at the La Silla Paranal Observatory under program ID195.B-0283. The data employed in this work can be downloaded from the ESO Archive Facility. F. N.-L. gratefully acknowledges the sponsorship provided by the Federal Ministry for Education and Research of Germany through the Alexander von Humboldt Foundation. F. N. acknowledges funding by grants PID2019-105552RB-C41 and MDM-2017-0737-19-3 Unidad de Excelencia "María de Maeztu". D. A. G. acknowledges support by STFC grant ST/T000244/1.

\end{acknowledgements}

\bibliographystyle{aa}
\bibliography{../../../../BibGC}
\end{document}